\documentclass[%
twocolumn,
 reprint,
superscriptaddress,
showkeys,
amsmath,amssymb,
]{revtex4-1}

\usepackage[mathlines,modulo,pagewise]{lineno}

\usepackage{graphicx}
\usepackage{bm}
\usepackage[dvipsnames]{xcolor}
\usepackage{lettrine}

\begin{document}
\preprint{APS/123-QED}

\title{Symmetry breaking and chaos in evaporation driven Marangoni flows over bubbles}
\author{V. Chandran Suja}
\email{vinny@stanford.edu}
\affiliation{%
 Department of Chemical Engineering, Stanford, CA-94035\\
}
 \author{A. Hadidi}
 \affiliation{Department of Mechanical Engineering, UCLA, CA 90095}
  \author{A. Kannan}
  \affiliation{%
 Department of Chemical Engineering, Stanford, CA-94035\\
}

\author{G.G.Fuller}%
 \email{ggf@stanford.edu}
\affiliation{%
 Department of Chemical Engineering, Stanford, CA-94035\\
}%

\begin{abstract}
Understanding the dynamics of liquid films that make up bubbles is of practical and fundamental importance. Practically, this understanding is crucial for tuning bubble stability, while fundamentally, thin films are an excellent platform to study 2D flows.  Here we study the spatiotemporal film thickness dynamics of bubbles subjected to evaporation driven Marangoni flows. Initially, we demonstrate how bubble stability can be dramatically tuned with the help of evaporation driven flows. Subsequently, we reveal that the spatial symmetry of thickness profiles evolves non-monotonically with the volatile species concentration, with profiles being axisymmetric at the two extremes in concentration. At $50\%$ concentration, spatial symmetry breaks down and thickness fluctuations are chaotic everywhere in space, with the fluctuation statistics becoming spatially invariant and ergodic. For these cases, the power spectrum of thickness fluctuations follow the Kolmogorov $-5/3$ scaling - a first such demonstration for forcing by evaporation. These results along with the reported setup provide an excellent framework to further investigate 2D chaotic flows.


\end{abstract}

\pacs{Valid PACS appear here}
\maketitle

\null\newpage
\null\newpage

\begin{figure*}[!th]
\includegraphics[width=\linewidth]{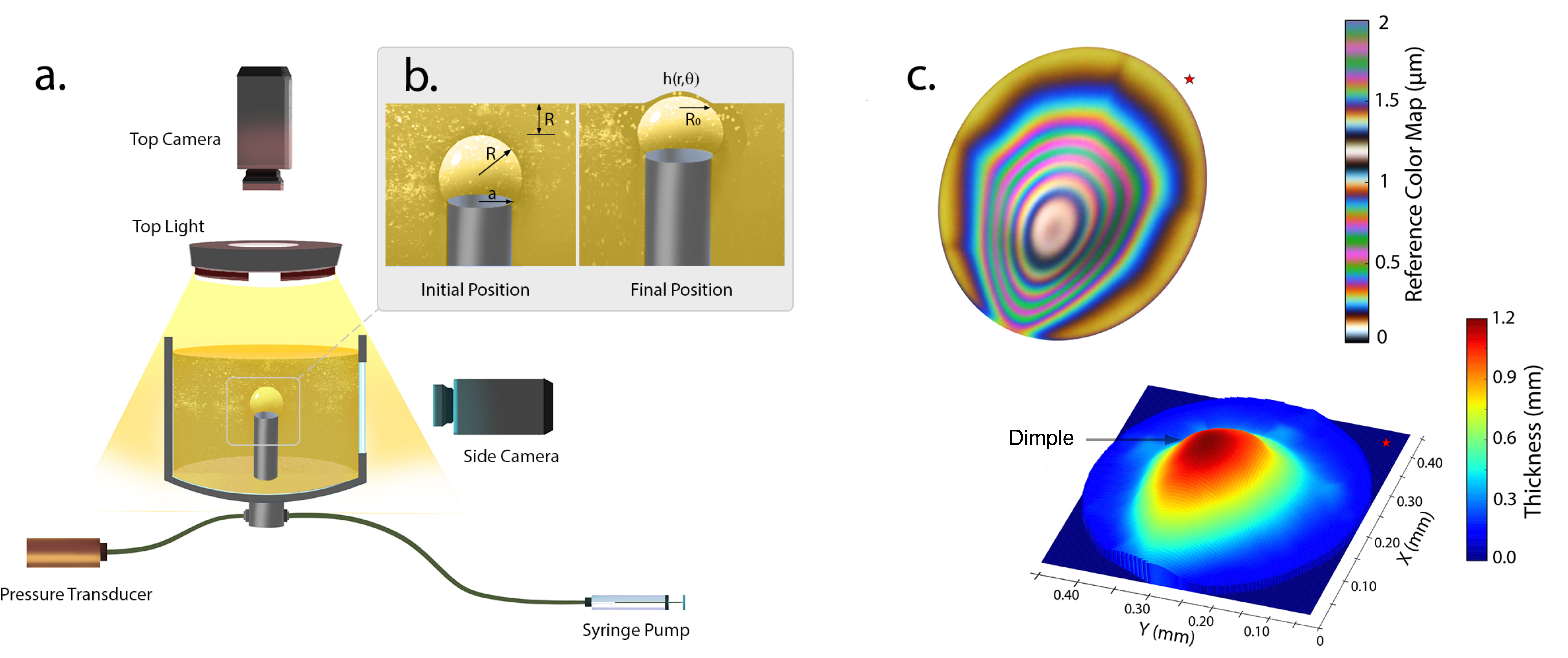} \caption{Schematic of the Dynamic Fluid-Film Interferometer (DFI) used to perform the single bubble experiments and a typical interferogram obtained from the experiments. ({\bf a.}) The setup with the labeled components. ({\bf b.}) The initial and final positions of the bubble during an experiment. Here $R$ is the radius of curvature of the bubble, $h(r,\theta)$ is the film thickness as a function of the radial position ($r$) and angular position $\theta$, and $R_0$ is the radial extend of the film visible on the interferogram. ({\bf c.}) An interferogram showing the asymmetric dissipation of a dimple (the thick region in the film) following its instability. The adjacent tile  shows physical film thickness corresponding to the interferogram. The fluctuations in the thickness visible along the barrier ring, the thinnest region in the periphery of the film, are a precursor to the instability of the dimple.}\label{fig:ExperimentalSetup}
\end{figure*}


\lettrine[lines=2]{\textcolor{Orange}B}{ubbles} are ubiquitous. They are an essential part of many foods, beverages, health care products and industrial processes \cite{frostad2016dynamic, garrett2016science, pugh1996foaming}. In other cases such as during lubrication, textile manufacturing and paper production, bubbles are an unwanted byproduct \cite{binks2010non,chandran2016impact,lantz2017filtration,friberg2010foams}. In all these cases, it is essential to control the life time of these bubbles. This is possible through a thorough understanding of the mechanisms that determine bubble stability.   

Evaporation is one such mechanism that affects bubble coalescence. The mass loss due to evaporation can have three different consequences on bubble stability. First, and the most obvious one, is the linear decay of the film surrounding the bubble due to the loss of mass.  This mechanism works to reduce  bubble stability and dictates bubble life times in many aqueous systems \cite{li2012effect,miguet2020stability}. Second, the temperature changes that accompany evaporation drive thermocapillary Marangoni flows. This mechanism usually works to enhance bubble stability and dictates bubble life times in highly volatile single component liquids with low specific heats \cite{menesses2019surfactant,neville1937effect}. Third, in multicomponent liquids, the species concentration changes that accompany evaporation drive solutocapillary Marangoni flows. These flows can dramatically alter bubble stability and is discussed in detail in section \textbf{\textit{Bubble stability and solutocapillary flows}}. This mechanism dictates bubble life times in many liquid mixtures and is prevalent in non-aqueous systems such as lubricant oils \cite{suja2018evaporation,shi2020oscillatory}.

In addition to their importance in bubble stability, the characteristics of Marangoni flows and thin film dynamics have broad implications in many other fields. Marangoni flows play an important role in dictating film spreading \cite{kavehpour2002evaporatively,nikolov2002superspreading}, fluid mixing \cite{scriven1960marangoni,schatz2001experiments} and breakup \cite{stone1994dynamics,thoroddsen2006crown,kamat2018role}. As the strength of the  Marangoni flows increase, the resulting dynamics can become chaotic, often leading to enhanced mixing or vigorous emulsification at fluid interfaces \cite{sternling1959interfacial}. Chaotic flows in thin films are also widely studied, primarily as a practical experimental platform to study the characteristics of 2D turbulence \cite{gharib1989liquid,greffier2002thickness,rivera1998turbulence,paret1997experimental,boffetta2012two,jullien2000experimental,hidema2010image}. Thin soap films flowing past obstacles have been extensively used to study decaying 2D turbulence \cite{greffier2002thickness,rivera1998turbulence,hidema2010image}, while electromagnetically actuated films have been used study forced turbulence \cite{paret1997experimental,boffetta2012two,jullien2000experimental}. These efforts have been successful in experimentally proving a number of theoretical \cite{kraichnan1967inertial} and numerical \cite{xiao2009physical} predictions relevant to 2D turbulence including the inverse energy cascade \cite{paret1997experimental}, the Kolmogorov \cite{greffier2002thickness} and Batchelor statistics \cite{jullien2000experimental}.  Further, experimental investigations have also uncovered links between the statistics of the thickness fluctuation in turbulent thin films and the statistics of passive scalars in turbulent 3D flows \cite{paret1997experimental}.  

In this manuscript, we experimentally study the spatial and temporal dynamics of film thickness profiles over bubbles in binary liquid mixtures subjected to evaporation driven Marangoni flows.  We initially demonstrate how bubble stability can be dramatically tuned with the help of evaporation driven flows. Next, we probe the spatial structure of the film thickness over time and establish surprising links between the volatile species concentration and the spatial symmetry of the film. Through the analysis of the temporal structure of the film thickness fluctuations over space, we reveal a non-monotonic dependence between volatile species concentration and the spatial prevalence of fluctuation stochasticity. At $50\%$  volatile species concentration, we show that these thickness fluctuations are chaotic everywhere in space, with the fluctuation statistics becoming spatially invariant and  ergodic. For these cases, the power spectrum of thickness fluctuations forced by evaporation  are shown to follow the Kolmogorov $-5/3$ scaling - a first such demonstration for forcing by evaporation. Finally, a simple analysis is developed to account for the surprising behaviors observed with changes in volatile species concentration. These insights into evaporation driven Marangoni flows along with the reported experimental setup and protocols also serve as an excellent platform to further our understanding of externally forced 2D chaotic flows.             







\section*{Results}
\subsection*{System}
We utilize single bubble experiments to probe the dynamics of evaporation driven Marangoni flows over bubbles. These experiments are performed in the so-called Dynamic Fluid-Film Inteferometer, hereon referred to as the DFI (Fig.\ref{fig:ExperimentalSetup}a). The details regarding its construction are mentioned elsewhere \cite{frostad2016dynamic} and in references therein. At the start of every experiment reported here, $5$ to $6\; ml$ of the binary liquid mixture is filled in the DFI chamber. A bubble of $1.2 \pm 0.15\; \mu l $ is then made on the tip of a $16$  gauge capillary (Inner Dia: $1.194 \pm 0.038\; mm)$. Subsequently, the bubble (radius $R \approx 0.7 \; mm$) is moved to the air-liquid interface at a constant velocity until it penetrates a fixed distance into the initially flat surface. The bubble is held at this final position (see Fig. \ref{fig:ExperimentalSetup}b) and the spatiotemporal evolution of the bubble wall thickness is measured using an optical arrangement (see details in \textit{\textbf{Methods}}). The internal bubble pressure is also measured, and the bubble coalescence time is identified accurately to $0.05\;s$ by tracking the dramatic changes in this pressure upon coalescence. The binary mixtures reported here are blends of silicone oils having $50\;cSt$ silicone oil as the non-volatile component and each of $0.65$, $1$, $1.5$ and $2\; cSt$ at six different concentrations as the volatile component (see \textit{\textbf{SI Materials, Table S1}} for properties).


\subsection*{Bubble stability and solutocapillary flows}
\begin{figure}[!th]
\includegraphics[width=\linewidth]{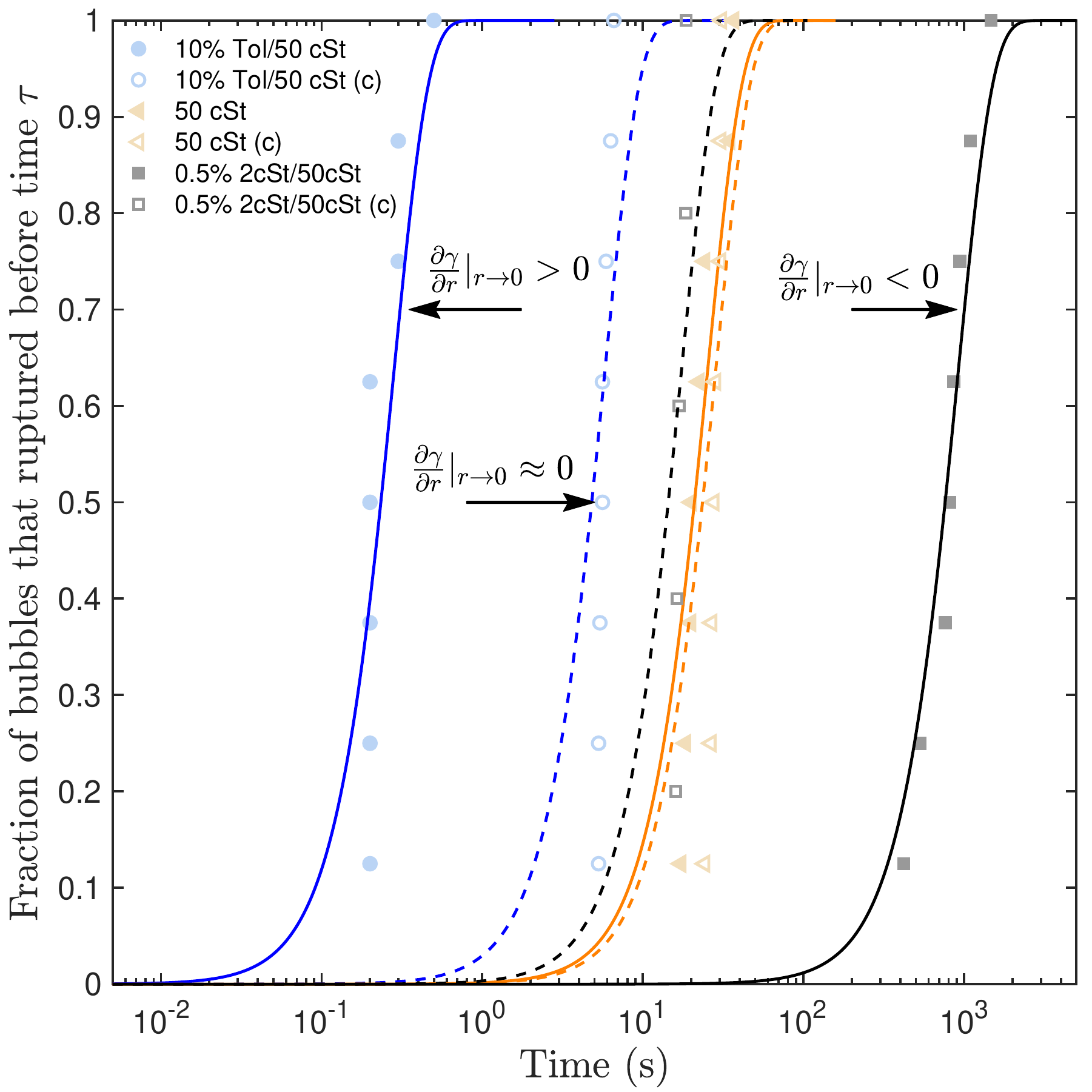} \caption{Influence of the radial direction of evaporation driven solutocapillary flows on bubble stability. The curves, referred to as the cumulative coalescence time curves \cite{suja2018evaporation}, are obtained by fitting experimentally measured coalescence times (symbols) to cumulative distribution functions of Rayleigh distributions.  The dotted curves correspond to cases when evaporation is suppressed, denoted by (c) in the legend. In these cases, there are no gradients in surface tension ($\left. \frac{\partial \gamma}{\partial r}\right|_{r\rightarrow 0} \approx 0 $) and the bubble stability approaches that of the pure liquid ($50 cSt$ silicone oil in this case). When the induced surface tension gradient is positive ($\left. \frac{\partial \gamma}{\partial r}\right|_{r\rightarrow 0} > 0 $), the curves shift to the left (spanning small values of time), indicating a marked decrease in bubble stability. This is illustrated by probing the stability of bubbles in a $10\%$ by
volume mixture of Toluene (abbreviated as Tol in the legend)
in $50\;cSt$ silicone oil. Toluene has a higher surface tension than $50\;cSt$ silicone oil, and as a result, the evaporation of toluene lowers the surface tension at the bubble apex.  On the other hand, when $\left. \frac{\partial \gamma}{\partial r}\right|_{r\rightarrow 0} < 0 $, the curves shift to the right, indicating a dramatic increase in bubble stability. This is illustrated by probing the stability of bubbles in a $0.5\%$ by
volume mixture of $2\;cSt$ silicone oil in $50\;cSt$ silicone oil. $2\;cSt$ oil has a lower surface tension than $50\;cSt$ silicone oil, and as a result the evaporation of $2\;cSt$ silicone oil elevates the surface tension at the bubble apex.}\label{fig:solutocapillaryBubblestability}
\end{figure}

A practical motivation for this study is the dramatic effects of evaporation on bubble lifetimes and the need to control it in many commonly used liquid mixtures (eg. in the lubricant industry\cite{suja2018evaporation}). In Fig.\ref{fig:solutocapillaryBubblestability}, we demonstrate using cumulative coalescence time curves \cite{suja2018evaporation,suja2020foam}, how the radial direction of evaporation driven Marangoni flows (here on solutocapillary flows) influences bubble stability. The cumulative coalescence time curves are a convenient way to rank bubble stability, and are obtained by fitting experimentally measured coalescence times (symbols) to cumulative distribution functions of Rayleigh distributions (see \textit{\textbf{SI Materials}} for details). Here we test three systems with and without evaporation. First, pure non-volatile $50\;cSt$ serves as a control system and establishes bubble stability in the natural absence of evaporation. In this case, there are negligible solutocapillary flows owing to the absence of any induced surface tension gradients $\left(\left. \frac{\partial \gamma}{\partial r}\right|_{r\rightarrow 0} \approx 0 \right)$. Here $\gamma$ is the local surface tension, and $r$ the radial co-ordinate measured from bubble apex (see Fig.\ref{fig:ExperimentalSetup}b). Second, a $10\%$ by volume mixture of Toluene (abbreviated as Tol in Fig.\ref{fig:solutocapillaryBubblestability}) in $50\; cSt$ silicone oil serves to study bubble stability when $\left. \frac{\partial \gamma}{\partial r}\right|_{r\rightarrow 0} >0$. Evaporation of toluene leads to changes in toluene concentration, with the local changes in concentration scaling inversely with the local film thickness (see \textit{\textbf{SI Materials}}). The film thickness is smaller near the apex of the bubble as compared to bulk, resulting in evaporation causing a larger depression in concentration near the bubble apex. As toluene has a higher surface tension than $50\;cSt$ silicone oil, there is a larger depression in surface tension at the bubble apex as toluene evaporates, resulting in a positive surface tension gradient that drive solutocapillary flows away from the bubble apex. These flows dramatically destabilize the bubbles as is evident by the shifting of the curves to the left, relative to the control case without evaporation. Conversely, a $0.5\%$ by volume mixture of $2\;cSt$ silicone oil in $50\; cSt$ silicone oil presents the case where $\left. \frac{\partial \gamma}{\partial r}\right|_{r\rightarrow 0} < 0 $. In this case, evaporation of $2\;cSt$ oil (surface tension lower than $50\; cSt$ silicone oil) leads to an elevation in surface tension at the bubble apex, driving solutocapillary flows towards the top of the bubble. As a consequence, the bubble stability is dramatically enhanced as evident by the shifting of the curves to the right, relative to the control case without evaporation. When evaporation is suppressed by covering the chamber (dashed curves), bubbles in all the binary mixtures behave similar to those in pure non-volatile liquids as a result of negligible induced surface tension gradients.


\begin{figure*}[!tp]
\includegraphics[width=0.9\linewidth]{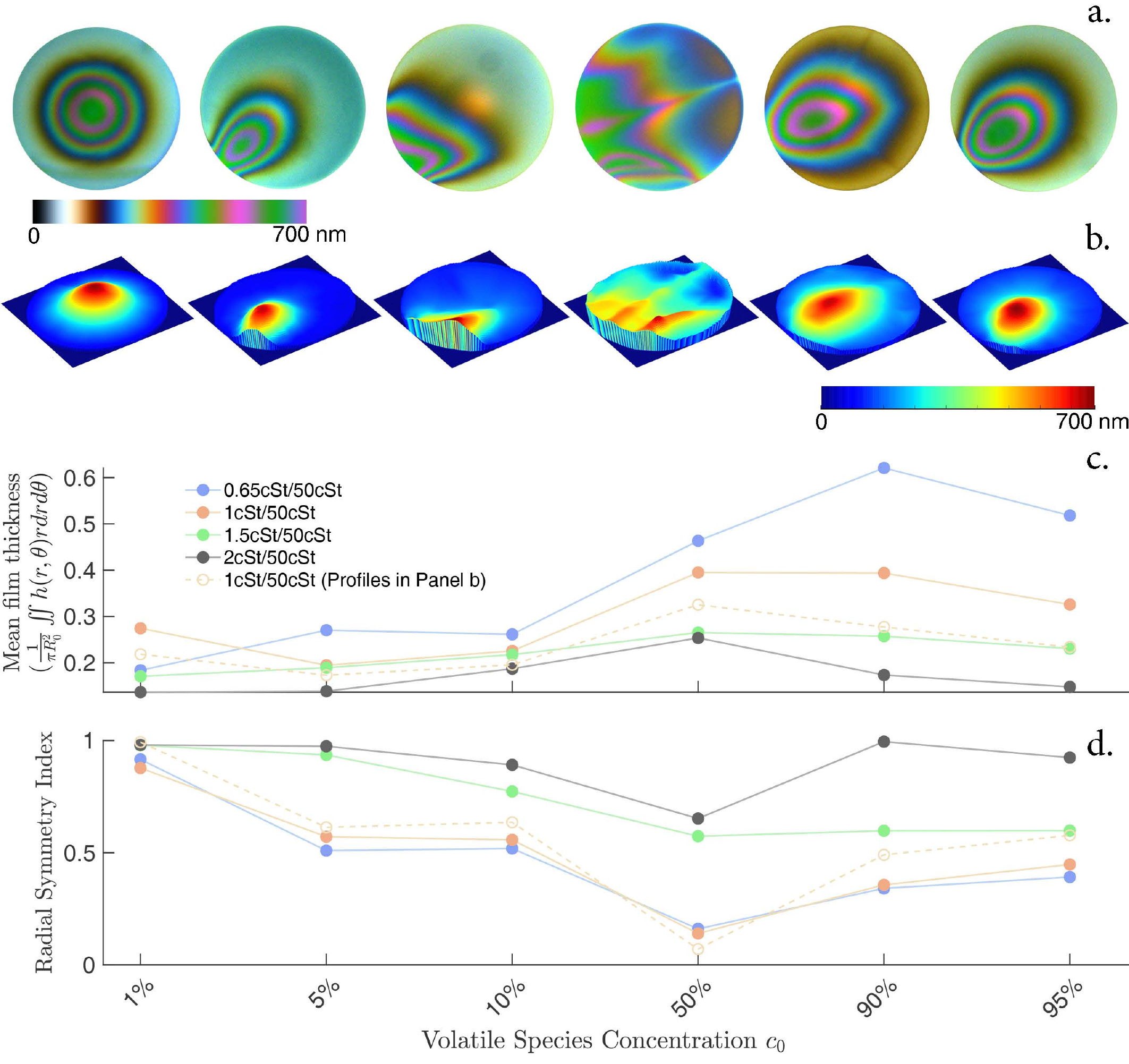} \caption{The spatial structure of film thickness as a function of volatile species volume concentration and chemistry. {\bf a.} Representative interferograms from bubbles in binary mixtures of $1 \;cSt$ and $50\; cSt$ silicone oils. For purposes of comparison, these interferograms were chosen from instances when the maximum film thickness was approximately $700\; nm$. {\bf b.} The physical film thickness profiles corresponding to the interferograms. {\bf c.} The time averaged mean film thickness ($\frac{1}{T}\int_0^T \hat{h}dt$) as a function of volatile species concentration and chemistry (temporal thickness data collected over 30 s in intervals of 0.5 s). Here $\hat{h} = (\pi R_0^2)^{-1} \iint h(r,\theta,t)rdrd\theta$ (see Fig.\ref{fig:ExperimentalSetup}b for notation). The time averaged mean film thickness of the films in general increase with volatility of the volatile species. On the other hand, time averaged thickness has a non-monotonic dependence on volatile species concentration.The hollow symbols indicate the values of the metric corresponding to the physical thickness profile shown in the panel above. {\bf d.} The time averaged radial symmetry index (RSI) as a function of volatile species concentration and chemistry. The radial symmetry of the film thickness profiles also changes non-monotonically with species concentration irrespective of the species chemistry.  }\label{fig:structure}
\end{figure*}

These results have three important takeaways. First, thermocapillary flows have negligible influence on the reported results. Evaporation always leads to a depression of surface temperature and consequently a positive surface tension gradient. If thermocapillary gradients were dominant, bubbles in the tested binary mixtures would been have stabilized irrespective of the solutocapillary gradients.  Second, binary mixtures that generate positive surface tension gradients can be used as intrinsic defoamers to destabilize bubbles \cite{garrett2016science}. Finally, binary mixtures that generate negative surface tension gradients stabilize bubbles, and by extension can stabilize foam. Since foam control is important in many applications involving liquid mixtures \cite{suja2018evaporation}, it is important to further investigate the later scenario over a broad range of volatile species concentration. Further, the competition of curvature induced flows and solutocapillary flows lead to complex dynamics \cite{velev1993spontaneous,suja2018evaporation} that merit detailed investigation from a fundamental perspective.  This will be done in the subsequent sections. 

\subsection*{Structure and symmetry of film thickness profiles}

The bubble film thickness profiles offer important insights into the dynamics at the bubble wall and are seen to be significantly influenced by the volatile species concentration and its properties. \textit{\textbf{Supplementary Video V1}} offers a vivid example of these dynamics for the case of $1\;cSt$ silicone oil mixed into $50\; cSt$ silicone oil. In figure \ref{fig:structure}, we explore in detail the spatial structure of film thickness as a function of volatile species volume concentration and chemistry. Panels \textit{\textbf{a}} and \textit{\textbf{b}} respectively show representative interferograms and the corresponding physical film profiles over bubbles in various binary mixtures of $1\; cSt$ and $50\; cSt$ silicone oil. For purposes of comparison,
these interferograms were chosen from instances when the maximum film thickness was approximately 700 nm. Qualitatively, a couple of interesting observations stand out. Firstly, as the concentration of the volatile $1\;cSt$ silicone oil increases, the average film thickness increases till the concentration reaches about $50\%$. A further increase in $1\;cSt$ silicone oil concentration results in a decrease in the mean film thickness.  Secondly, a similar non-monotonic dependence is observed between the volatile species concentration and the radial symmetry. 

The above observations are quantified in the remaining panels of Fig.\ref{fig:structure}. Panel \textit{\textbf{c}} shows the time averaged mean film thickness ($\hat{\bar{h}} = \frac{1}{T}\int_0^T \hat{h}dt$) as a function of volatile species chemistry and concentration (temporal thickness data collected over 30 s in intervals of 0.5 s). Here $\hat{h} = (\pi R_0^2)^{-1} \iint h(r,\theta)rdrd\theta$ is the mean (space averaged) film thickness (see Fig.\ref{fig:ExperimentalSetup}b for notation). For a given concentration, the time averaged mean film thickness increases with the volatile species concentration. This is a consequence of the stabilizing Marangoni flux scaling proportionally to the solvent volatility. On the other hand, for a given volatile species, the time averaged mean film thickness first increases and then decreases as the volatile species concentration is increased. The rationale for this non-monotonic dependence is also related to the variations in the strength of the Marangoni flux, and is discussed in detail in section \textit{\textbf{Perturbation sensitivity}} and \textit{\textbf{SI Materials}}. 

Panel \textit{\textbf{d}} reports the time averaged Radial Symmetry Index (RSI) that quantifies the radial symmetry of the film profiles as viewed from the apex of the dimple (equivalently the location of the maximum film thickness - see Fig.\ref{fig:ExperimentalSetup}c). Physically, RSI is a measure of the variance in film thickness at every radial location and ranges from 0 indicating large variance or no symmetry to 1 indicating little variance or complete symmetry (see \textit{\textbf{SI Materials}} for details). The RSI starts close to 1 for all the systems at low concentration of the volatile species. As the volatile species concentration increases, the RSI decreases indicating a decrease in radial symmetry. Further increase in volatile species concentration (beyond $50\%$) leads to an increase in RSI indicating a recovery in radial symmetry. The solvent volatility significantly influences the spatial symmetry, with RSI scaling inversely with the species volatility at any given concentration. It is also worth noting for binary mixtures comprised of highly volatile $0.65\;cSt$ and $1\;cSt$ silicone oils, RSI approaches zero at $50\%$ species concentration, indicating a total loss of spatial symmetry. The rationale for the observed behavior in radial symmetry is related to the variation in the sensitivity of the system to ambient disturbances that alter the evaporation rate over the bubble, and is discussed in detail in section \textit{\textbf{Perturbation Sensitivity}}.

\subsection*{Structure and statistics of film thickness fluctuations}
\begin{figure*}[!tp]
\includegraphics[width=0.9\linewidth]{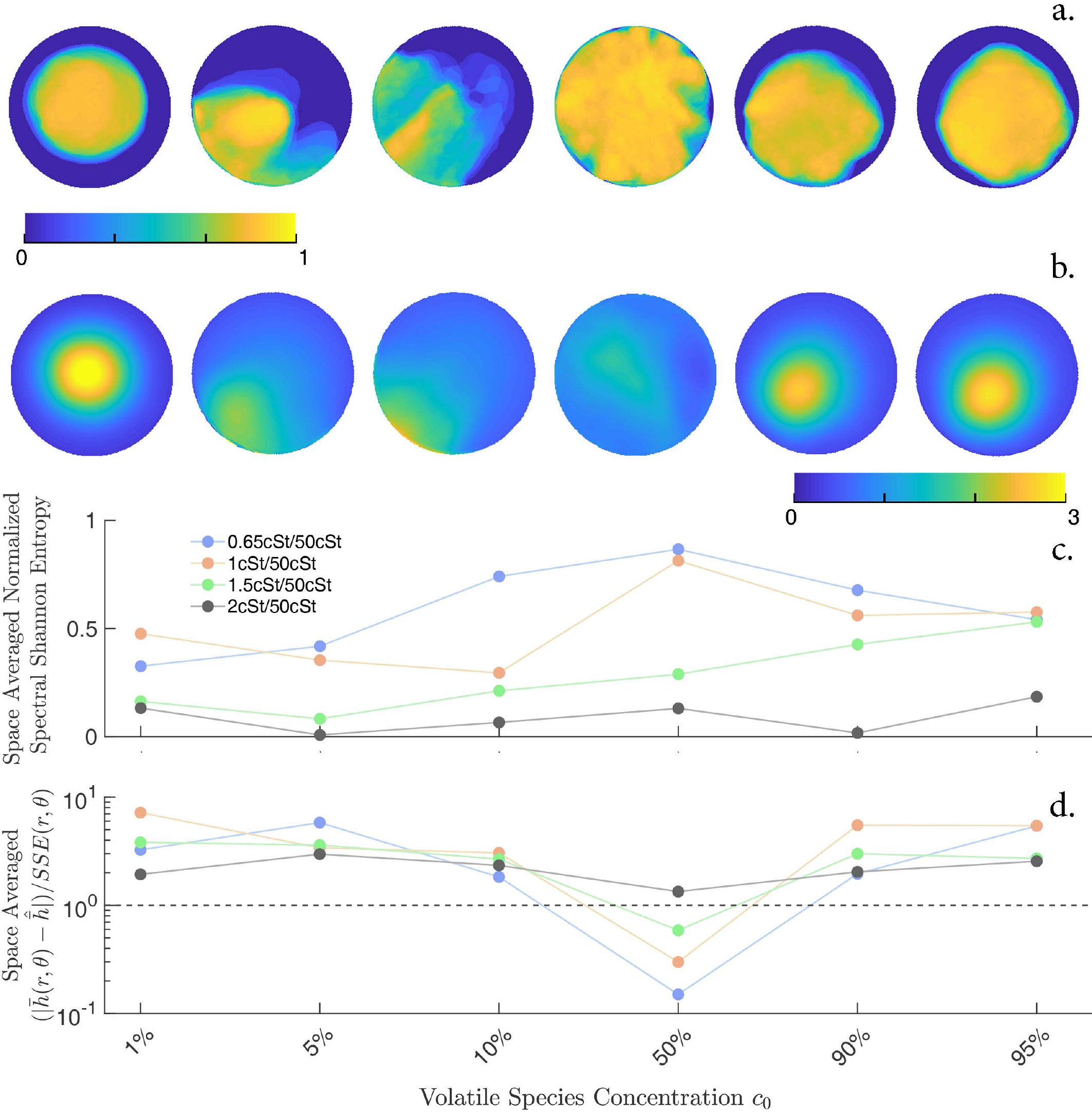} \caption{The structure of film thickness fluctuations as a function of volatile species volume concentration and chemistry. {\bf a.} Contours of normalized spectral Shannon entropy for bubbles in binary mixtures of $1\;cSt$ and $50\; cSt$ silicone oil. The normalized spectral Shannon entropy quantifies the stochasticity in film thickness fluctuations, with a value of 0 signifying complete order (or predictability) and a value of 1 signifying complete disorder. {\bf b.} Contours of $\bar{h}(r,\theta)/\hat{\bar{h}}$ in binary mixtures of $1\;cSt$ and $50\; cSt$ silicone oil. Here $\bar{h}(r,\theta)=\frac{1}{T}\int_0^T h(r,\theta,t) dt$ is the time averaged film thickness and  $\hat{\bar{h}} = (\pi R_0^2)^{-1} \iint \bar{h}(r,\theta)rdrd\theta$ is the time averaged mean film thickness (the same quantity reported in Fig.\ref{fig:structure}). It is interesting to note that $\bar{h}(r,\theta)/\hat{\bar{h}}$ approaches 1 everywhere in space for the case with $50\%$ concentration of the volatile species. {\bf c.} The space averaged normalized spectral Shannon entropy (SSE) as function of volatile species volume concentration and chemistry. At fixed concentration,  SSE scales with species volatility and for the cases of $0.65\; cSt$ and $1\; cSt$ the SSE is very high (approaches 1) at $50\;\%$ concentration. {\bf d.} The space average of $\left(|\bar{h}(r,\theta) - \hat{\bar{h}}|\right)/SSE(r,\theta)$ as function of volatile species volume concentration and chemistry. $\left(|\bar{h}(r,\theta) - \hat{\bar{h}}|\right)/SSE(r,\theta)$ is physically related to the variance of $SSE^{-1}$ weighted time averaged film thickness distribution in space. The weighting with the inverse of SSE ensures low values of this metric are not a result of static regions. Further, regularization (by adding a small number, in this case by adding 0.01 to both the numerator and denominator) ensures a value of 1 for this metric for the case of a purely static film. This limit is indicated by the hashed line. It is interesting to note that this quantity approaches zero for the highly volatile $0.65\; cSt$ and $1\; cSt$ silicone oils. }\label{fig:fluctuations}
\end{figure*}

The fluctuations in bubble film thickness are also influenced by the volatile species concentration and its properties \cite{suja2018evaporation,shi2020oscillatory}. In Fig.\ref{fig:fluctuations} we probe this dependence by exploring the space averaged temporal structure of film thickness fluctuations as a function of volatile species volume fraction and chemistry. Panel {\bf a} shows contours of the normalized spectral Shannon entropy for various binary mixtures of $1\;cSt$ and $50\:cSt$ silicone oil. The normalized spectral Shannon entropy (SSE) is calculated as $\Sigma_{i=0}^{N} p_i \log_N pi$, where the microstate probability $p_i$ is identically set equal to the normalized amplitude of the $N$ Fourier modes of the signal (see \textit{\textbf{SI Materials}} for details). The SSE quantifies the stochasticity in temporal fluctuations, with a value of 0 signifying complete order (or predictability) and a value of 1 signifying complete disorder. At low volatile species concentration, high entropy is localized to the dimple (see Fig.\ref{fig:structure} for the physical thickness profile). Similar aperiodic oscillations of the dimple have previous been observed for binary systems with low concentration of volatile species \cite{suja2018evaporation,shi2020oscillatory}. It is interesting to note that despite the localized stochasticity in film thickness fluctuations, there is spatial order for these films as seen from Fig.\ref{fig:structure}c, thus physically indicating that these stochastic fluctuations have long range spatial correlations.  As the volatile species concentration increases, the stochastic oscillations become asymmetric, consistent with the asymmetric location of the dimple (Fig.\ref{fig:structure}b). Further increase in volatile species concentration up to $50\;\%$ results in the stochasticity in thickness fluctuations enveloping larger regions in space, with almost the complete bubble surface becoming stochastic at $50\;\%$. From Fig.\ref{fig:structure}c, we also know there is a concurrent break down in spatial order at this concentration. This implies that the long range spatial correlations in fluctuations that existed at lower concentrations are now absent. Further increase in volatile concentration beyond $50\;\%$, results in an increasing localization of the stochastic fluctuations, along with an increase in spatial order, implying a recovery of long range spatial correlation. 

Panel {\bf b} shows contours of $\bar{h}(r,\theta)/\hat{\bar{h}}$ in binary mixtures of $1\;cSt$ and $50\; cSt$ silicone oil. Here $\bar{h}(r,\theta)=\frac{1}{T}\int_0^T h(r,\theta,t) dt$ is the time averaged film thickness and  $\hat{\bar{h}} = (\pi R_0^2)^{-1} \iint \bar{h}(r,\theta)rdrd\theta$ is the time averaged mean film thickness (the same quantity reported in Fig.\ref{fig:structure}). Physically, these contours reveal the variance in the time averaged film thickness across space. At $1\%$ concentration of the volatile species, we see that time averaged film thickness is skewed in space, with the highest values of the metric overlapping with the apex of the dimple. As the volatile species concentration increases and the spatial order breaks down, the spatial variance decreases. It is interesting to note that $\bar{h}(r,\theta)/\hat{\bar{h}}$ approaches 1 almost everywhere in space for the case with $50\%$ concentration of the volatile species, despite the high stochasticity and lack of spatial order. Physically, this implies every point in space is sampling all possible thicknesses and no point in space is special. This is often referred to as spatial invariance and is a feature of fully developed chaos. At still higher concentrations, the metric starts deviating from 1 and high values are localized similar to the case at very low concentrations of the volatile species.   

Panel {\bf c} quantifies the observations in panel {\bf a} through a plot of the space averaged SSE.  At a fixed concentration, space averaged SSE scales with the species volatility, and is particularly high for the highly volatile $0.65\;cSt$ and $1\;cSt$ silicone oils at a concentration of $50\%$. For a fixed species, we observe the SSE to increase up to $50\%$  before decaying at higher concentration for top two volatile species. Panel {\bf d} quantifies the observations in panel {\bf b} through a plot of the space average of $(|\bar{h}(r,\theta) - \hat{\bar{h}}|)/SSE(r,\theta)$. This metric is physically related to the variance of $SSE^{-1}$ weighted time averaged film thickness distribution in space. The weighting by the inverse of SSE ensures low values of this metric are not a result of static regions. Further, regularization (by adding a small number, in this case by adding 0.01 to both the numerator and denominator) ensures this quantity has value of 1 for the case of a purely static film. This limit is indicated by the hashed line. For all species and for all concentrations (except $50\%$), the value of this quantity is above 1. At $50\%$ concentration,  $(|\bar{h}(r,\theta) - \hat{\bar{h}}|)/SSE(r,\theta)$ scales inversely with the species volatility. At this concentration, this metric approaches 0 for systems with the two most volatile silicone oils - $0.65\;cSt$ and $1\;cSt$. This clearly shows that these two systems display spatial invariance along with high temporal stochasticity - a feature of fully developed chaos.  

\begin{figure*}[!tp]
\includegraphics[width=\linewidth]{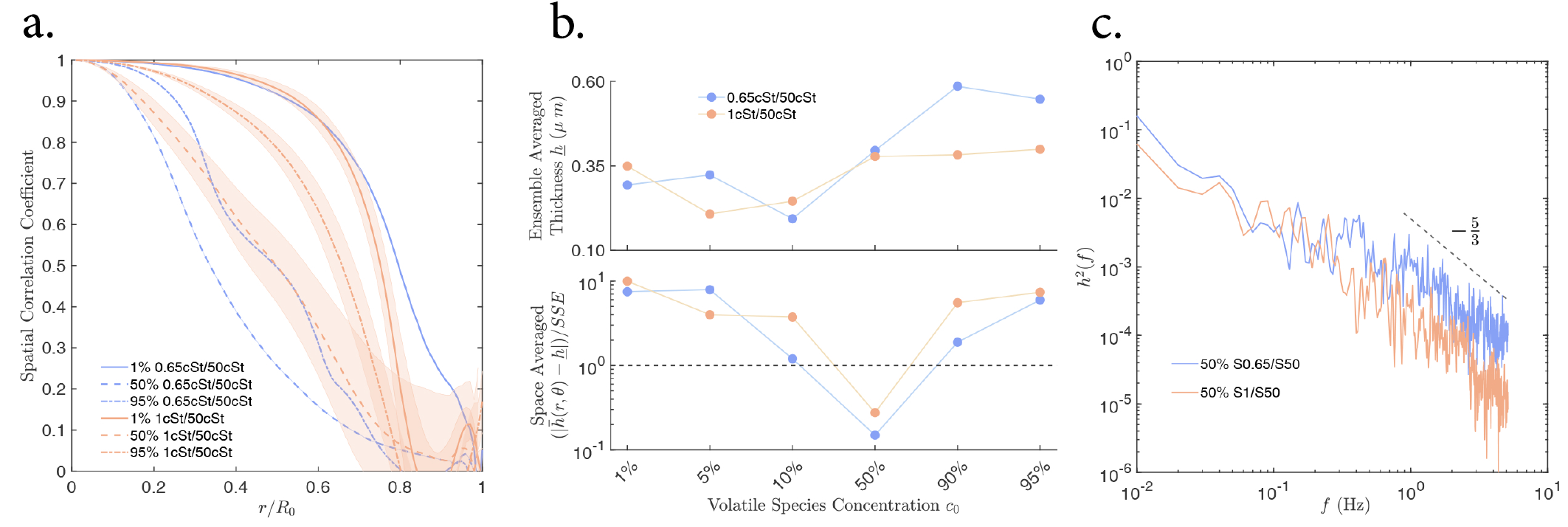} \caption{Features of chaotic (turbulent) flows observed in binary silicone oil mixtures of $0.65\;cSt$ in $50\; cSt$ and $1\;cSt$ in $50\; cSt$ at $50\%$ volatile species concentration. {\bf a.} The spatial correlation coefficient for thickness fluctuations as a function of the non-dimensional radial distance from the bubble center. The shaded regions (only shown for $1\;cSt$ in $50\; cSt$) indicates the standard deviation of the correlation coefficient. Clearly there is a sharp decline in the spatial correlation distance at any given radial location for the case with $50\%$ concentration as compared to either very low or very high concentration of the volatile species. The increase in the standard deviation is a direct consequence of the emergence of spatial disorder at $50\%$ volatile species concentration. {\bf b.} Top panel shows the ensemble mean of the spatial thickness measured at a fixed time ($t = 5s$) from the start of the experiment across five different experiments. Bottom panel shows the spatial average of the absolute deviation between the temporal and ensemble average of the film thickness. The weighting with the inverse of SSE ensures low values of this metric are not a result of static regions (see Fig.\ref{fig:fluctuations} caption). At $50\%$ concentrations, the absolute deviation between the temporal and ensemble average of the film thickness approaches zero, signifying film thickness fluctuations are ergodic. {\bf c.} The power spectrum of the thickness fluctuations measured at the bubble apex for the case with $50\%$ concentration of the volatile species. The energy spectrum obeys the famed Kolmogorov $-5/3$\textsuperscript{rd} decay \cite{greffier2002thickness}. Further, the energy at a given frequency scales with the volatility of the evaporating species.   }\label{fig:chaos}
\end{figure*}

\subsection*{Chaos: Turbulent thin film flows}
As seen from the previous sections, at $50\%$ concentration, binary silicone oil mixtures of $0.65\;cSt$ in $50\; cSt$ and $1\;cSt$ in $50\; cSt$ display traits consistent with chaotic (turbulent) flows. In Fig.\ref{fig:chaos}, we probe the the spatiotemporal evolution of the film thickness in the above two systems and highlight the features of the observed chaotic flows. Panel {\bf a} shows the spatial correlation coefficient for thickness fluctuations as a function of the non-dimensional radial distance
from the bubble apex for three different volatile species concentration in mixtures of $0.65\;cSt$ in $50\; cSt$ and $1\;cSt$ in $50\; cSt$. The shaded regions (only shown for $1\;cSt$ in $50\;cSt$) indicate the standard deviation of the correlation
coefficient. As expected, the spatial correlation extends over large radial distances when the species concentration is very low ($1\%$) or very high ($95\%$). For these cases,  there is a dramatic decay in the spatial correlation beyond a critical radial distance corresponding to the extend of the dimple (Fig.\ref{fig:structure}). On the other hand at $50\%$ species concentration, there is a dramatic decay in spatial correlation at smaller radial distances. This indicates weak spatial correlation, which is a common feature of chaotic flows.  The widening of the standard
deviation (shaded region) at $50\%$ volatile species concentration is yet another indication of the emergence of spatial disorder. 

The top subplot in panel {\bf b} shows the ensemble average thickness as a function of volatile species concentration.  The ensemble average thickness is calculated by averaging the spatial thickness measured at a fixed time ($t = 5s$) from the start of the experiment across five different experiments. The bottom subplot in panel {\bf b} shows the SSE inverse weighted spatial average of the absolute difference between the temporal and ensemble average thickness. The weighting with the inverse of SSE ensures low values of this metric are not a result of static regions (see Fig.\ref{fig:fluctuations} caption and the previous section for details). This metric follows trends similar to that observed for the spatial variance of time averaged thickness (Fig.\ref{fig:fluctuations}d) and tends to zero at $50\%$ concentration. Physically, this implies that everywhere in space the temporal mean approaches the ensemble mean. In other words thickness fluctuations are ergodic - another characteristic of fully developed chaotic flows.  

Panel {\bf c} shows the power spectrum of the thickness fluctuations measured at the bubble apex for the case with $50\%$ concentration of the volatile
species. Two interesting observations stand out. First, the power spectrum of the thickness fluctuations obey a power law scaling that closely resembles the Kolmogorov $-5/3$\textsuperscript{rd} decay. Such a scaling for thickness fluctuations has been previously observed in turbulent planar soap films flowing past a periodic array of cylinders \cite{greffier2002thickness}. However, this observation has never been reported over curved bubbles nor for the case where evaporation drives the turbulent flows in the films. Physically, the agreement with the Kolmogorov scaling reveals that the statistics of film thickness fluctuations in chaotic evaporation driven flows are also similar to those of a passive scalar field in a turbulent flow. Second, the energy contained in fluctuations scales with the volatility of the evaporating species, as observed by the larger amplitude of $h^2(f)$ at a fixed frequency for the more volatile $0.65\;cSt$ oil. Physically, this reaffirms the role of evaporation in actively driving the thickness fluctuations. Further, this observation also implies that the intensity of the chaotic flows (turbulence intensity) can be modulated by changing the species volatility.

\subsection*{Perturbation sensitivity}
\begin{figure}[!h]
\includegraphics[width=0.9\linewidth]{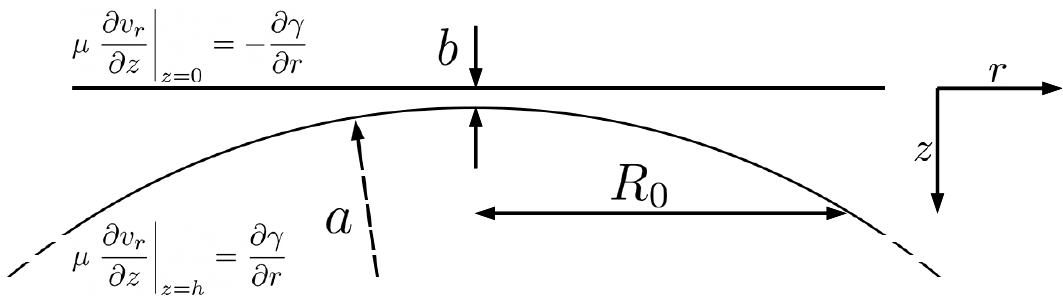} \caption{Schematic of a simplified scenario similar to the reported experiments. $b$ is the separation distance of the bubble apex from the planar interface and is  comparable to the apex thickness of bubbles ($\sim \mathcal{O}(10^{-6})m$. $R_0$ is the radial extend of the domain of interest and is analogous to the radial extend of the interferograms. $a$ is the radius of the bubble. Here $b\ll R_0 \ll a$.  The origin of the cylindrical coordinate system is on the planar interface and coincides with the radial location of bubble apex.}\label{fig:simpleSchematic}
\end{figure}

A detailed understanding of the observed dynamics requires the coupled solution of an advection-diffusion equation with a source term for species transport and the lubrication equations for the flow field \cite{rodriguez2019evaporation,shi2020oscillatory}. As we are dealing with free films whose shapes are not known apriori, analytical solutions are intractable.  We can readily obtain physical insights into the problem by making assumptions on the shape of the boundaries. One such physically realizable scenario is shown in Fig.\ref{fig:simpleSchematic}. 

For illustration, let us consider how the radial velocities evolve with the different parameters in the problem.  Integrating the radial momentum equation in the lubrication limit along the $z$ direction and applying the tangential stress boundary conditions gives us the following relationship between the depth averaged radial velocity and the evaporation induced surface tension gradients (see \textit{\textbf{SI Materials}} for details), 
\begin{equation}\label{eq.radvelgrad}
\left<v_r\right> = \frac{1}{h}\int_0^h v_r dz  \propto  \frac{h}{\mu}\frac{\partial \gamma}{\partial r}
\end{equation}
Assuming instantaneous diffusion in $z$ direction, negligible diffusion in the $r$ direction and negligible convective species transport, we obtain the following expression for the radial gradient in surface tension (see \textit{\textbf{SI Materials}} for details),
\begin{equation}\label{eq.TensionGrad}
        \frac{\partial \gamma}{\partial r} =  \frac{\partial\gamma}{\partial c} \frac{\partial h}{\partial r}\frac{  c_0(1-c_0) }{h^2} E t
\end{equation}
Here the local surface tension $\gamma =\gamma(c(r,\theta,t),\Delta \gamma,\gamma_{nv})$. $\Delta \gamma = \gamma_{v} - \gamma_{nv}$ is the equilibrium surface tension differential between the volatile and the non-volatile species, and $c$ is the local species concentration.  $\frac{\partial\gamma}{\partial c}$ has the same sign of $\Delta \gamma$, and for binary liquid mixtures obeying the linear mixing rule, $\frac{\partial\gamma}{\partial c} = \Delta \gamma$. At short times $\frac{\partial\gamma}{\partial c}\approx \frac{\partial\gamma}{\partial c_0}$. $E$ is the evaporation flux of the pure volatile species per unit area. For mixtures, we assume the evaporation rates are proportional to the area of the solution surface occupied by the volatile species, which in turn scales with $c_0$, the initial species volume concentration. Assumptions on diffusion are valid as gradients in $z$ are of the order $\mathcal{O}(\frac{1}{h}\sim \frac{1}{b})$, while the gradients in $r$ are of the order $\mathcal{O}(\frac{\partial h}{\partial r} \sim \frac{R_0}{a})$. On the other hand, assuming negligible convective species transport will only strictly hold at short times. Nevertheless, Eq.\ref{eq.radvelgrad} along with Eq.\ref{eq.TensionGrad} provide physical insights into the observed dynamics. 

There are four important takeaways from the above equations. First, we observe that the depth averaged radial velocity $\left<v_r\right>$ scales with $E$ and is zero when evaporation rate $E$ is $0$. Hence these dynamics require at least one volatile component. Second, the direction of $\left<v_r\right>$ is dictated by the sign of $\Delta \gamma$ as we saw in section \textit{\textbf{Bubble stability and solutocapillary flows}}. Third, we observe $\left<v_r\right>$ depends on the geometry of the film. Spatial gradients in thickness are necessary to generate these flows. Further, the flow strength scales inversely with the film thickness. Finally, we also observe $\left<v_r\right>$ scales non-linearly with the initial species concentration as $c_0(1-c_0)$.  

The fluctuations in film thickness are driven by the radial flows. Hence, the observed break down of symmetry as well as the  stochasticity in the film thickness fluctuations can be rationalized by analyzing the sensitivity of the $\left<v_r\right>$ to ambient perturbations. Among the quantities that influence $\left<v_r\right>$, the evaporation flux $E$ is strongly modulated by the ambient perturbations. Mathematically, this sensitivity is simply given by the magnitude of $\frac{\partial\left<v_r\right>}{\partial E}$. For a given species chemistry, $\mu$ and $\frac{\partial\gamma}{\partial c}\approx \frac{\partial\gamma}{\partial c_0}$ are implicit functions of concentration. Hence we have $\frac{\partial\left<v_r\right>}{\partial E} \propto \frac{\partial\gamma}{\partial c_0} \frac{c_0(1-c_0)}{\mu}$. Clearly this quantity approaches zero both at very low and high concentrations, and has a single maxima (both $\frac{\partial\gamma}{\partial c_0}$ and ${\mu}$ monotonically depend on concentration). As far as the explicit contribution of species concentration is concerned, $\frac{\partial\left<v_r\right>}{\partial E}$ clearly reaches a maximum when $c_0 = 0.5$. Thus the system is insensitive to perturbations at very low and very high concentrations, while the sensitivity reaches a maximum at an intermediate concentration.  Similar arguments can be extended to understand azimuthal velocity variations and by extension azimuthal fluctuations in thickness.




\subsection*{Discussion and Outlook}

In summary, the bubble stability and the spatiotemporal dynamics of film thickness profiles over bubbles subjected to evaporation driven Marangoni flows are dictated by the volatile species properties and concentration. Bubble stability is dramatically influenced by radial direction of the Marangoni flows, and can be tuned by choosing an appropriate combination of volatile and non-volatile species. When $\Delta \gamma$, the equilibrium surface tension differential between the volatile and non-volatile species is positive, the flows are destabilizing, while when $\Delta \gamma < 0$, the flows stabilize the bubble.   

For latter case, consistent with the previous reports \cite{chandran2016impact,shi2020oscillatory}, at low volatile species concentration $c_0$, the film thickness profiles are spatially symmetric. Despite spatial order, the temporal fluctuations are chaotic with long range spatial correlation as observed from the contours of the normalized spectral Shannon Entropy of the system.  As the volatile species concentration increases, the spatial thickness distribution becomes increasingly asymmetric. Concurrently, the stochasticity in film thickness fluctuations envelop larger regions in space.  At $50\%$ species concentration, there is a complete loss of spatial symmetry in the film thickness distribution and the thickness fluctuations become stochastic almost everywhere in space, along with a dramatic drop in the spatial correlation length. A further increase in species concentration beyond $50\%$, results in a gradual recovery in the spatial symmetry of the thickness distribution. Analogous to the behavior at low concentration, the stochastic fluctuations in the film thickness become localized in space and the spatial correlation length increases. 

The similarity in system dynamics at the two extremes in concentration, as well as the pervasive spatial disorder and stochasticity in film thickness fluctuations at $50\%$ concentration, can be rationalized by analyzing the sensitivity of flows inside the bubble wall to ambient fluctuations. Since ambient fluctuations primarily influence the dynamics through modulating the evaporation rate $E$, a physical understanding of the system sensitivity can be obtained by computing the sensitivity of flow velocities in the film to variations in $E$, $\frac{\partial \left< v_r \right>}{\partial E}$. At short times, $\frac{\partial \left< v_r \right>}{\partial E}  \propto \frac{\partial\gamma}{\partial c_0} \frac{c_0 (1-c_0)}{\mu }$. Clearly, the sensitivity vanishes both when $c_0 \rightarrow 0$ and when $c_0 \rightarrow 1$. As a result, there is spatial symmetry and thickness fluctuations have long range spatial correlation at the two extremes in volatile species. At intermediate concentrations, the system is very sensitive to ambient fluctuations, explaining the spatial disorder and short range spatial correlation in thickness fluctuations.  

At $50\%$ species concentration, spatio-temporal film thickness fluctuations in binary mixtures of $0.65\;cSt$ in $50\;cSt$ and $1\;cSt$ in $50\;cSt$ silicone oils display characteristics of chaotic (turbulent) flows. First, the temporal mean of thickness fluctuations are almost identical everywhere in space. In other words, temporal mean of film thickness fluctuations are spatially invariant. Second, the ensemble mean of thickness fluctuations across different experimental realizations approach the temporal mean of thickness fluctuations everywhere in space. In other words, film thickness fluctuations are ergodic. Third, the power spectra of the thickness fluctuations obey a power law scaling that closely resembles the Kolmogorov $-5/3$\textsuperscript{rd} decay. Physically, this  power law scaling implies that the statistics of film thickness fluctuations in chaotic evaporation driven flows are similar to those of a passive scalar field in a turbulent flow. Such a scaling has been previously observed in planar soap films \cite{greffier2002thickness}, and has never been reported over curved bubbles or for the case where evaporation drives the turbulent flows.

There remains several opportunities for future work that would offer important extensions to the present study. First and foremost, a detailed understanding of the dimple instability would be of fundamental interest. A good understanding of the stability of dynamically formed dimples and the related stability criteria is available in literature \cite{joye1994asymmetric,joye1996numerical}. However, the stability criteria for the evaporation driven spontaneously formed dimples are yet to be developed. Second, evaporation driven 2D turbulence is a little explored area of research. The experimental setup and protocols described in this manuscript provide a convenient platform to further research in this area. A number of exciting research directions may be pursued in this area using this setup that includes but are not limited to investigating features and energy cascades of 2D forced turbulence and the features of velocity fields in 2D evaporation driven turbulence.       



\section*{Methods}
\subsection*{Materials}
Silicone oils (Shin-Etsu Chemical Co., Ltd, Japan) are used as the model systems to study the kinematics and dynamics of evaporation-driven spontaneous dimpling.  For this study, we use a non-volatile oil with viscosity of $50 \; cSt$. To make the binary mixtures reported in this study, the non-volatile oil is each mixed with one of the four volatile oils having viscosities of $0.65\; cSt$, $1\;cSt$, $1.5\;cSt$ and $2\;cSt$, at volume fractions. The physical properties including the evaporation rates are reported in Supplementary Materials (Section 1).


\subsection*{Single bubble measurements}
Single bubble coalescence experiments are conducted using the Dynamic Fluid-Film Interferometer(DFI); the specific details regarding its construction are mentioned elsewhere \cite{frostad2016dynamic} and in references therein. At the start of every single bubble experiment, 5 to 6 mL of the lubricant is placed into the DFI chamber. A bubble of $1.2 \pm \; 0.15 \mu l$ is made at the tip of a standard 16 guage capillary (OD: $1.651\pm0.013 \;mm$, ID: $1.194\pm0.038 \;mm$). The bubble size is chosen to be as close as possible to the bubble size having the largest bubble number density in a freshly formed foam \cite{deane2002scale}, and at the same time large enough to avoid instabilities associated with manipulating small bubbles on capillaries \cite{chandran2016impact}. After the bubble is created on the capillary, the chamber is moved down (with bubble remaining stationary) at a constant velocity of $0.15 \;mm/s$ until the bubble is one radius away from the oil-air interface. This is the initial state of the system before all the experiments (Fig.\ref{fig:ExperimentalSetup}b).

At this point the experiment starts with the pressure transducer measuring the pressure inside the bubble at $20\;Hz$.  After the pressure is monitored for 10 seconds (to make sure there are no fluctuations in the size of the bubble), the oil-air interface is lowered a distance of 1.5 times the bubble radius from its initial position and held at that final position. This final position is comparable to the equilibrium position attained by free bubble through the balance of buoyancy and capillary forces. Simultaneously, the top camera records the evolution of the film of liquid between the bubble and the lubricant-air interface. As the film drains and its thickness becomes comparable to the wavelength of light, interference patterns are seen by the top camera. Finally the experiment ends as the film ruptures and the bubble coalesces at some critical film thickness.  The film thickness is obtained by mapping the colors in the recorded interference patterns to physical thickness using the classical light intensity - film thickness relations \cite{sheludko1967thin} assuming  homogeneous and non dispersive films. A Python $2.7$ based software was developed in-house \cite{frostad2016dynamic} to aid thickness mapping and visualizing the thickness profiles. 

\subsection*{Evaporation measurements}
A Mettler-Toledo (Model: AB135S, OH, USA) weighing balance, accurate to $10 \; \mu g$, was to quantify the volatilities of the pure silicone oils and silicone oil mixtures. $4\; ml$ of the test liquid is filled in a chamber identical to the one used to perform the single bubble experiments. The assembly is then left on the weighing balance, with the mass-loss data collected at $1\;Hz$ over the course of 10000 seconds. The reported evaporation rates (Table) are calculated by extracting the slope of the best fit line to the mass-loss data, and subsequently dividing this slope ($g/s$) by the surface area available for evaporation. 

\subsection*{Acknowledgements}
We thank Aparna Vijayan for assiting with data analysis, Benjamin Chadwick for assistance with running the experiments and Prem Sai for the graphical illustrations in the manuscript.  

\subsection*{Author Contribution}
VCS conceived the study, performed the experiments, developed the analysis and drafted the manuscript. AH performed the experiments. AK characterized the tested binary mixtures. GGF supervised the study. 

\bibliography{References.bib}
\end{document}